\documentclass{PoS}

\pdfoutput=1

\usepackage{amsmath}

\newcommand{\lst}[2]{#1^3\!\!\times\!#2}

\newcommand{\pbp}{\langle\bar\psi\psi\rangle}

\title{Evidence for a First Order, Finite Temperature Phase Transition
  in 8 Flavor QCD}

\ShortTitle{Evidence for a First Order, Finite Temperature Phase Transition
  in 8 Flavor QCD}

\author{\speaker{Xiao-Yong Jin}\\
        Department of Physics, Columbia University, New York, NY 10027, USA\\
        E-mail: \email{xj2106@columbia.edu}}

\author{Robert D. Mawhinney\\
        Department of Physics, Columbia University, New York, NY 10027, USA\\
        E-mail: \email{rdm@physics.columbia.edu}}

\abstract{

  As part of our ongoing investigations of QCD with many flavors of
  quarks, here we report on studies of the finite temperature phase
  transition for eight-flavor QCD with the DBW2 gauge action and
  na\"ive staggered fermions.  We find a clear first order phase
  transition between the chirally asymmetric phase at zero temperature
  and the chirally symmetric phase at finite temperature, signaled by
  a two-state signal for $\pbp$ at a non-zero temperature. We see this
  signal at a gauge coupling of $\beta=0.54$, where, to set the scale,
  the zero temperature value for $f_\pi$, in the chiral limit, is
  $0.06661(92)$.  This strong, first-order signal is seen for two
  different values of the quark mass, $m_q=0.007$ and $0.0195$, at
  $N_\tau=8$ and $6$ respectively.  Using $f_\pi(m_q)$ as the scale,
  the critical temperature is measured to be $T_c/f_\pi=1.638(93)$ at
  $m_\pi/f_\pi=3.329(30)$ for $m_q=0.007$, and $T_c/f_\pi=1.779(27)$
  at $m_\pi/f_\pi=4.093(15)$ for $m_q=0.0195$.  At a weaker coupling
  $\beta=0.56$, where at zero temperature and in the chiral limit we
  find $f_\pi=0.0312(10)$, the first order signal becomes numerically
  invisible to us for the $N_\tau \leq 14$ lattices we have
  investigated so far.

}

\FullConference{The XXVIII International Symposium on Lattice Filed Theory\\
		 June 14-19,2010\\
		 Villasimius, Sardinia Italy}

\begin{document}

\section{Introduction}

\begin{figure}
  \centering
  \includegraphics{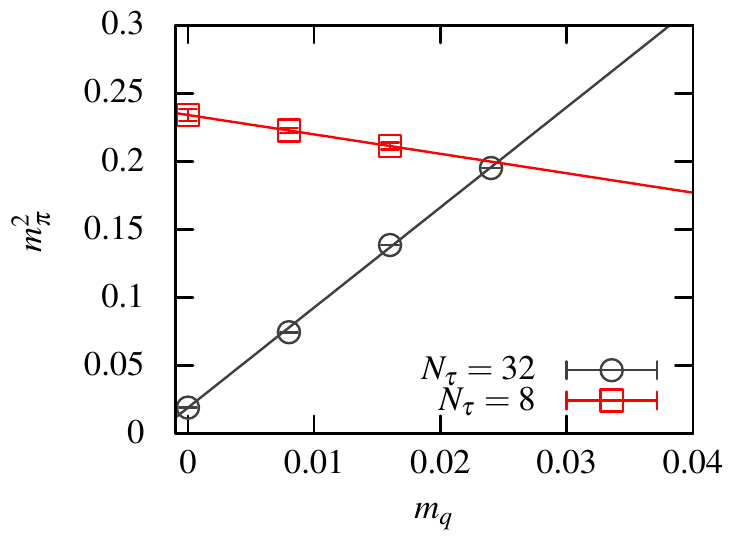}
  \includegraphics{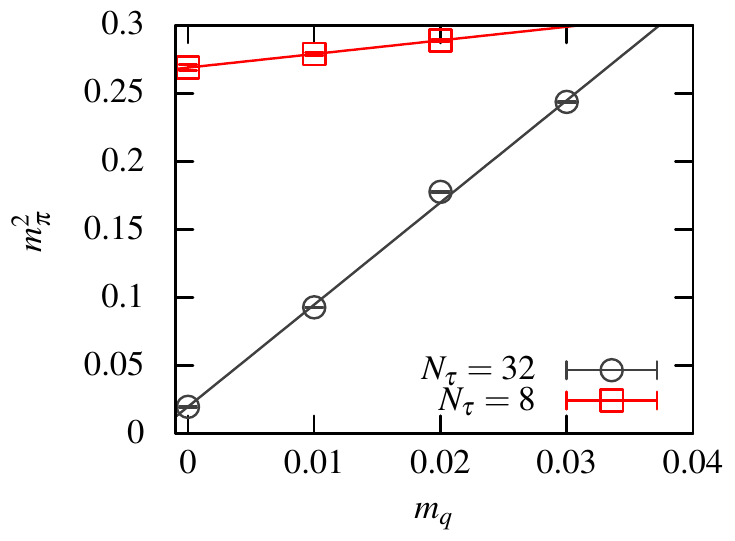}
  \caption{Pion screening masses at finite temperature ($N_\tau=8$)
    compared to pion masses at zero temperature ($N_\tau=32$).  The
    left panel shows 8 flavors at $\beta=0.56$; the right panel shows
    12 flavors at $\beta=0.49$.}
  \label{fig:mpi_ext_nt8}
\end{figure}

\begin{figure}
  \centering
  \includegraphics{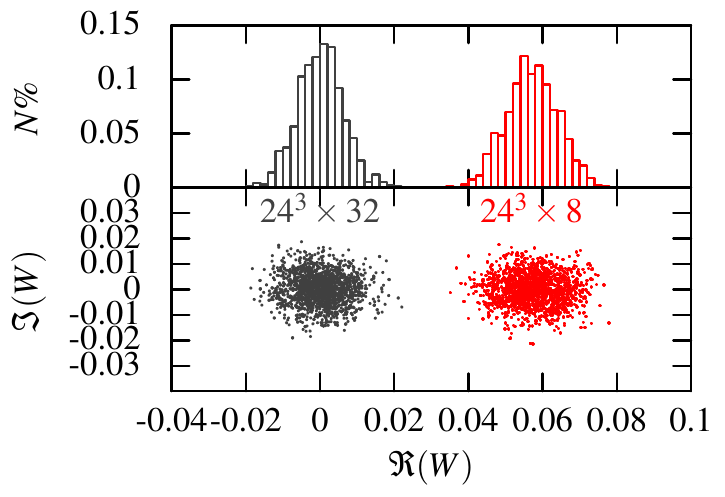}
  \includegraphics{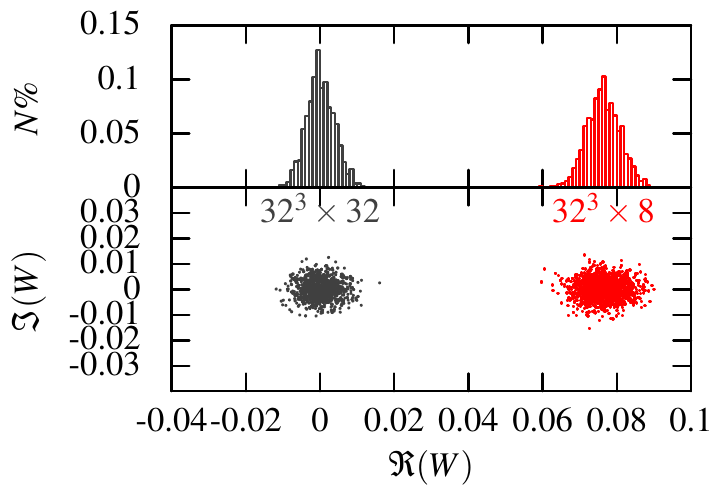}
  \caption{Values for Wilson loops along the temporal direction at finite
    temperature ($N_\tau=8$) compared to the ones at zero temperature
    ($N_\tau=32$).  The lower part of each figure shows a scatter
    plot of the real (horizontal axis) and imaginary parts (vertical
    axis) of the Wilson loops, while the upper part shows a histogram
    of the real part of the Wilson loops.  The histogram is normalized
    such that the height of each block represents the percentage
    of numbers in the bin.  The left figure shows 8 flavors with
    $m_q=0.008$ at $\beta=0.56$; the right shows 12 flavors with
    $m_q=0.01$ at $\beta=0.49$.  Lattice sizes are labeled in the
    figures.}
  \label{fig:wline_nt8}
\end{figure}

This work is a continuation of our ongoing studies of many flavor QCD,
as reported in previous conferences \cite{Jin:2008rc,Jin:2009mc}.  We
use the DBW2 gauge action and na\"ive staggered fermion action with
the exact RHMC algorithm.  We have studied a series of bare coupling
values and quark masses for both 8 and 12 flavors at zero temperature
($N_\tau=32$).  Various hadronic observables have been measured on the
lattices, to determine both the general lattice scale and the phase of
the theory.  Our conclusion is that QCD with both 8 and 12 flavors
exhibits properties that are consistent with spontaneous chiral
symmetry breaking at zero temperature.  We have also studied the
finite temperature behavior of the $N_f = 8$ and 12 flavor QCD at
$N_\tau=8$ \cite{Jin:2009mc}.  A notable result is the behavior of the
mass of the pion shown in Figure~\ref{fig:mpi_ext_nt8}.  At zero
temperature, we find $m_\pi^2 = B m_q$ for both 8 and 12 flavors,
which is the expected behavior for a Goldstone boson produced by
spontaneous chiral symmetry breaking.  Such behavior goes away at
finite temperature, where chiral symmetry is restored, as is clearly
evident in the figure.  Measurements of Wilson loops along the
temporal direction, shown in Figure~\ref{fig:wline_nt8}, support the
existence of a finite temperature phase by clearly showing a nonzero
value at $N_\tau=8$.

The chiral symmetry restoring finite temperature phase transition in
QCD is expected to be first order \cite{Pisarski:1983ms}, when the
number of massless fermions is equal to or larger than 3.  Since we
have established that chiral symmetry is spontaneously broken at zero
temperature and seen that it is restored at finite temperature, we
wanted to see if the transition is first order for a range of non-zero
quark masses accessible with current computer resources.  If the
transition is strongly first order, one should see the discontinuity
of the order parameter manifest itself in co-existing, metastable
evolutions of the lattice for precisely chosen parameters on large
volumes.  Non-zero fermion masses are expected to weaken any first
order signal, and, for large enough masses, the phase transition will
end in a second order critical point and then become a crossover.  If
a first order, finite temperature phase transition is found for
non-zero quark masses, one can be fairly confident that this will
persist as the quark masses go to zero.

In our previous, zero temperature work, we have seen a rapid change in
the lattice scale with the coupling for 8 flavors \cite{Jin:2008rc}
and a much more significant change for 12 flavors \cite{Jin:2009mc}.
To investigate the details of the finite temperature phase transition
and to interpret it with physical quantities, it is beneficial for us
to use $\beta$ values where we have zero temperature ensembles.  We
can then probe various discrete values of temperature or $N_\tau$ by
continuously tuning the bare quark mass as an input.  We chose $\pbp$
to be the order parameter and ran simulations starting with both
ordered and disordered gauge fields.  A discontinuity due to a first
order phase transition should result in two co-existing, metastable
thermalized lattice ensembles with the same set of input parameters
and having different values of $\pbp$.

\section{Finite temperature phase transition with 8 flavors}

\begin{figure}
  \centering
  \includegraphics{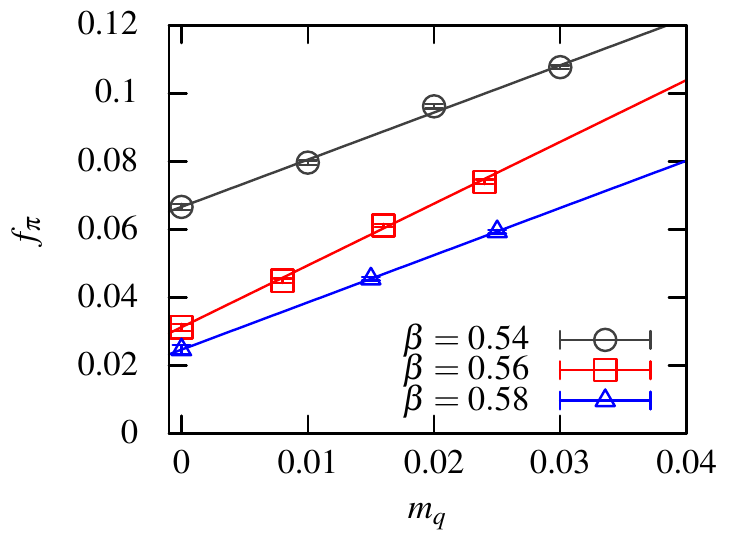}
  \caption{The pion decay constant, $f_\pi$, for 8 flavors at zero
    temperature, with a linear extrapolation to the chiral limit.}
  \label{fig:8f_fpi}
\end{figure}

In Figure~\ref{fig:8f_fpi} we show a linear extrapolation to the
chiral limit of $f_\pi$ for three $\beta$ values where we have
extensive zero temperature results for 8 flavors.  One sees that
between $\beta = 0.54$ and $0.56$, the lattice scale changes by almost
a factor of 3.  We expect the first order signal to be the strongest
at small masses and the strongest coupling, $\beta=0.54$, so we have
begun our simulations there.

\begin{figure}
  \centering
  \includegraphics{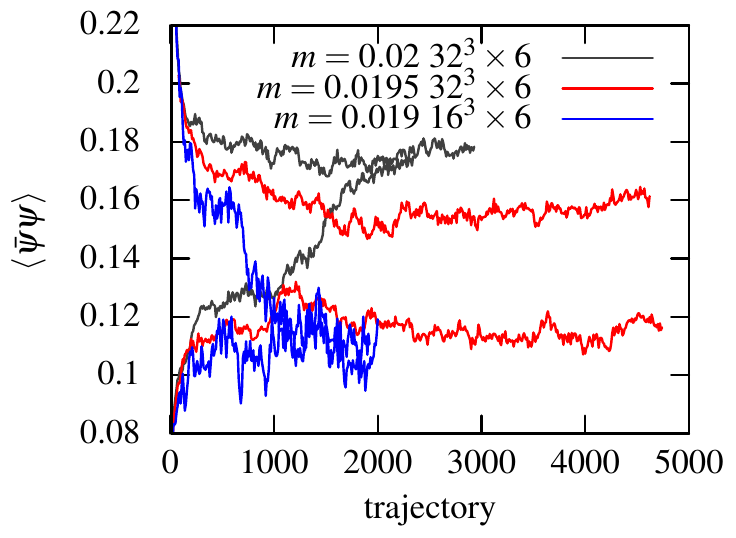}
  \includegraphics{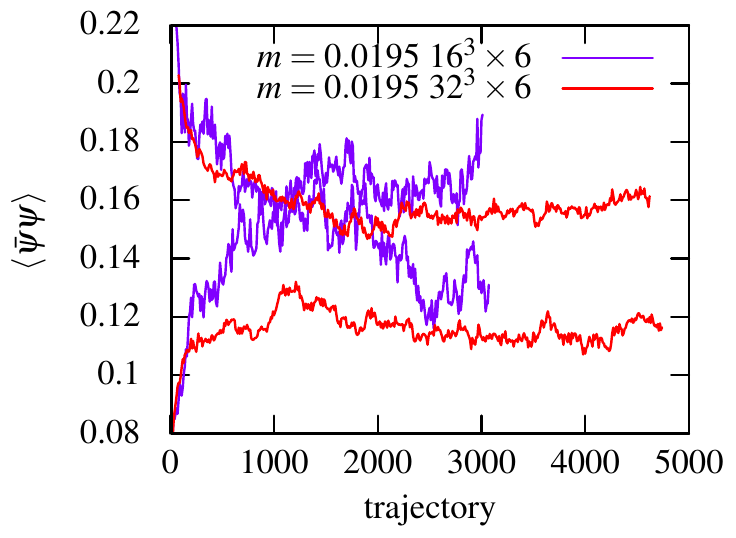}
  \caption{The evolution of $\pbp$, with both ordered (curves starting
    from bottom) and disordered starts (curves starting from top),
    using 8 flavors at $\beta=0.54$ and $N_\tau=6$.  The left panel
    shows the transition for 3 different masses.  The right panel
    shows 2 different lattice volumes at the transition mass,
    $m_q=0.0195$.}
  \label{fig:b054_pbp_nt6}
\end{figure}

Figure~\ref{fig:b054_pbp_nt6} shows the evolution of $\pbp$ at
$\beta=0.54$ and $N_\tau=6$.  By running with different quark masses,
a metastability for nearly 4000 trajectories is seen at $m_q=0.0195$
with a lattice size of $\lst{32}{6}$.  We find a discontinuity of
$\Delta\pbp = 0.0416(20)$ at this quark mass.  Metastability is very
sensitive to the quark mass; we have to tune the quark mass to within
$3\%$ to see two co-existing states.  A large spatial volume, $32^3$
is also required, as shown in the right panel of
Figure~\ref{fig:b054_pbp_nt6}, where the metastability is not apparent
on a $\lst{16}{6}$ volume.

\begin{figure}
  \centering
  \includegraphics{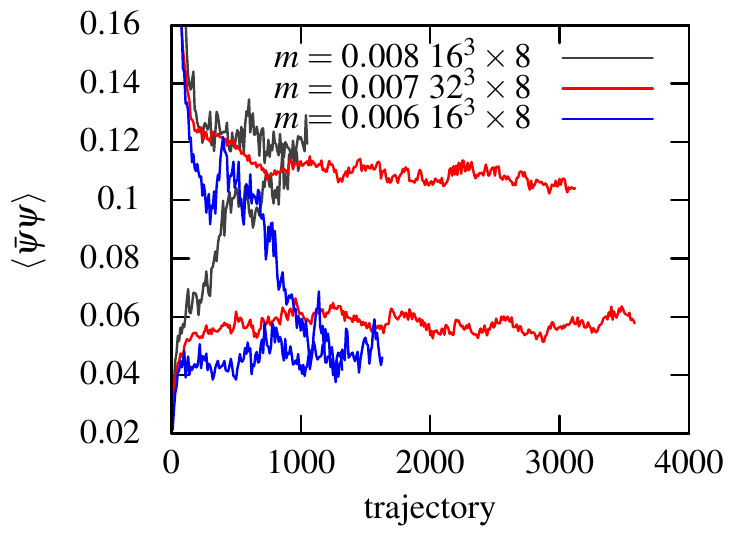}
  \includegraphics{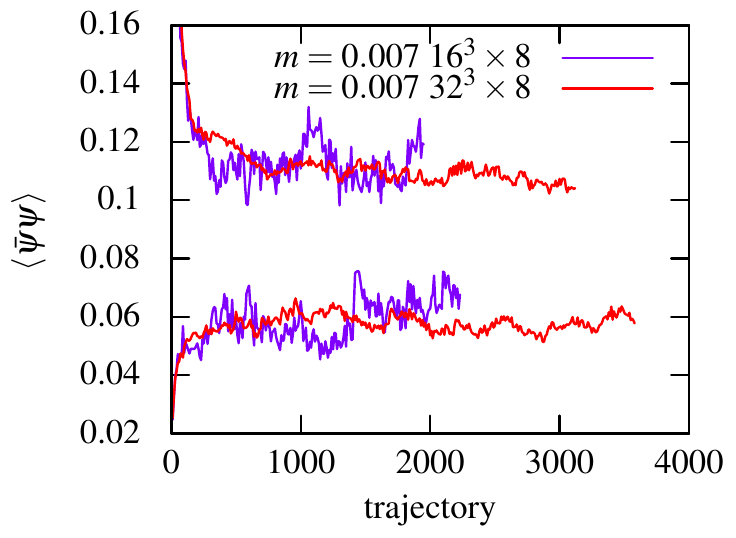}
  \caption{The evolution of $\pbp$ with both ordered (curve starting
    from bottom) and disordered starts (curve starting from top),
    using 8 flavors at $\beta=0.54$ and $N_\tau=8$.  The left panel
    shows the transition for 3 different masses.  The right panel
    shows 2 different lattice volumes at the transition mass,
    $m_q=0.007$.}
  \label{fig:b054_pbp_nt8}
\end{figure}

For $N_\tau=8$, shown in Figure~\ref{fig:b054_pbp_nt8}, metastability
is seen at a much smaller quark mass, $m_q=0.007$, for more than 2000
trajectories.  The first order signal is stronger, with a
discontinuity $\Delta\pbp=0.0508(17)$.  The quark mass is tuned within
$15\%$.  And both $16^3$ and $32^3$ volumes are clearly metastable.

\begin{figure}
  \centering
  \includegraphics{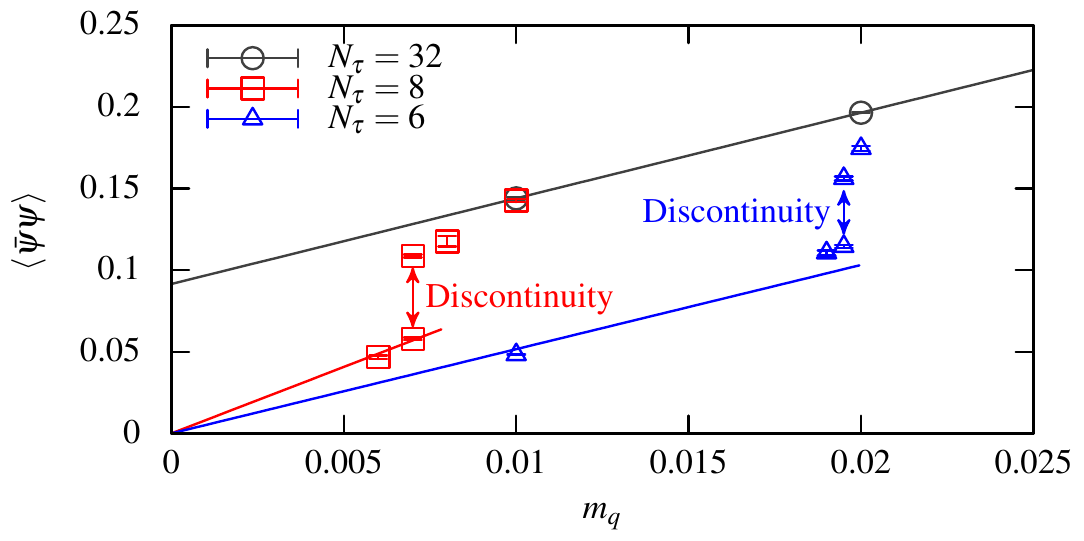}
  \includegraphics{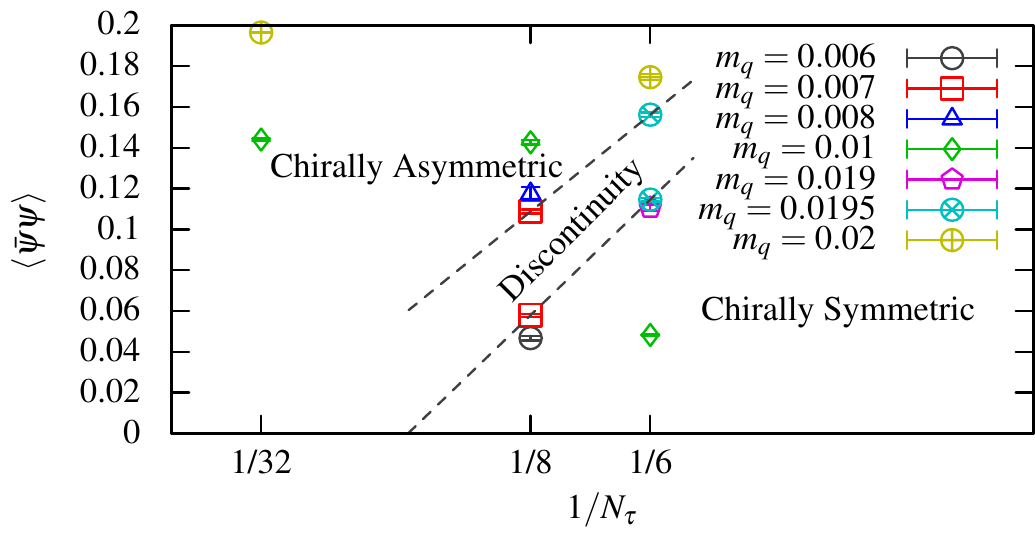}
  \caption{A diagram showing the values of $\pbp$ at $\beta=0.54$ with
    the discontinuity due to the first order phase transition
    explicitly shown.  The upper panel shows $\pbp$ versus $m_q$ and
    the lower panel shows $\pbp$ versus temperature ($1/N_\tau$).}
  \label{fig:b054_pbp}
\end{figure}

We can draw the chiral symmetry restoration transition diagram for 8
flavors at $\beta=0.54$ as shown in Figure~\ref{fig:b054_pbp}.  The
upper panel shows the value of $\pbp$ at different temperatures while
varying the quark mass.  The top line shows a linear extrapolation of
$\pbp$ at $N_\tau=32$ to the chiral limit.  The discontinuity
separates the chirally broken phase (upper points) and the chirally
symmetric phase (lower points).  The two lower lines, for $N_\tau=8$
and $N_\tau=6$, which have slopes of $8.14$ and $5.16$ respectively,
are linear fits to $\pbp$ in the chirally symmetric phase.  They are
forced to go to the origin, which is the expected behavior of $\pbp$
at finite temperature.  The fit at $N_\tau=6$ has a much larger
$\chi^2/\mbox{dof}$ (about 260) than the fit at $N_\tau=8$ (about 5).
This can be seen as a sign of weakened first order signal due to the
relatively large fermion mass.  The lower panel of
Figure~\ref{fig:b054_pbp} is another way of visualizing the phase
diagram.  The two dashed lines are linear fits to the metastable
signal of $\pbp$ at the edge of chirally broken phase (upper left
region), which has a slope of $1.14$, and the edge of chirally
symmetric phase (lower right region), which has a slope of $1.36$.  In
between is the inaccessible discontinuity region of $\pbp$.

\begin{table}
  \centering
  \begin{tabular}[c]{r|l|l}
    $m_q$ & 0.007 & 0.0195 \\
    $m_\pi(m_q)/f_\pi(m_q)$ & 3.329(30) & 4.093(15) \\
    \hline
    $T_c(m_q)/f_\pi(m_q)$ & 1.638(93) & 1.779(27)
  \end{tabular}
  \caption{The critical temperature, $T_c(m_q)$, of the first order,
    chiral symmetry restoring, finite temperature phase transition
    with 8 flavors, as observed at bare coupling $\beta=0.54$.}
  \label{tab:tc_fpi}
\end{table}

Table~\ref{tab:tc_fpi} displays the critical temperature, $T_c(m_q)$,
in units of $f_\pi(m_q)$.  The values of $f_\pi(m_q)$ and $m_\pi(m_q)$
come from na\"ive extrapolations of zero temperature results, where
$m_\pi^2 \propto m_q$ and $f_\pi \propto m_q$.  The quoted errors of
$m_\pi/f_\pi$ are only statistical errors.  The errors quoted for
$T_c/f_\pi$ uses error propagation from the error of the linear
extrapolation of $f_\pi$ and the estimated error of $T_c$ that is
described in the following.  First, we can estimate the error in our
determination of $m_q$ at the transition point from the range of quark
masses which either do or do not show metastability.  This gives us a
quark mass uncertainty of $0.001$ for $N_\tau=8$ and $0.0005$ for
$N_\tau=6$.  At the edge of the chirally symmetric phase, using the
two fits in the top panel of Figure~\ref{fig:b054_pbp}, this
uncertainty in $m_q$ can be translated to an uncertainty in $\pbp$ of
$0.00814$ and $0.00258$ at $N_\tau=8$ and $6$ respectively.  These
values can be further converted to an uncertainty on $T_c$ by the
relation of $\pbp$ and $T_c$ at the edge of chirally symmetric phase
which is represented by the lower dashed line in the lower panel of
Figure~\ref{fig:b054_pbp}.  This procedure results in estimated errors
on $T_c$ of $0.00599$ and $0.00190$ at $N_\tau=8$ and $6$
respectively.

\begin{figure}
  \centering
  \includegraphics{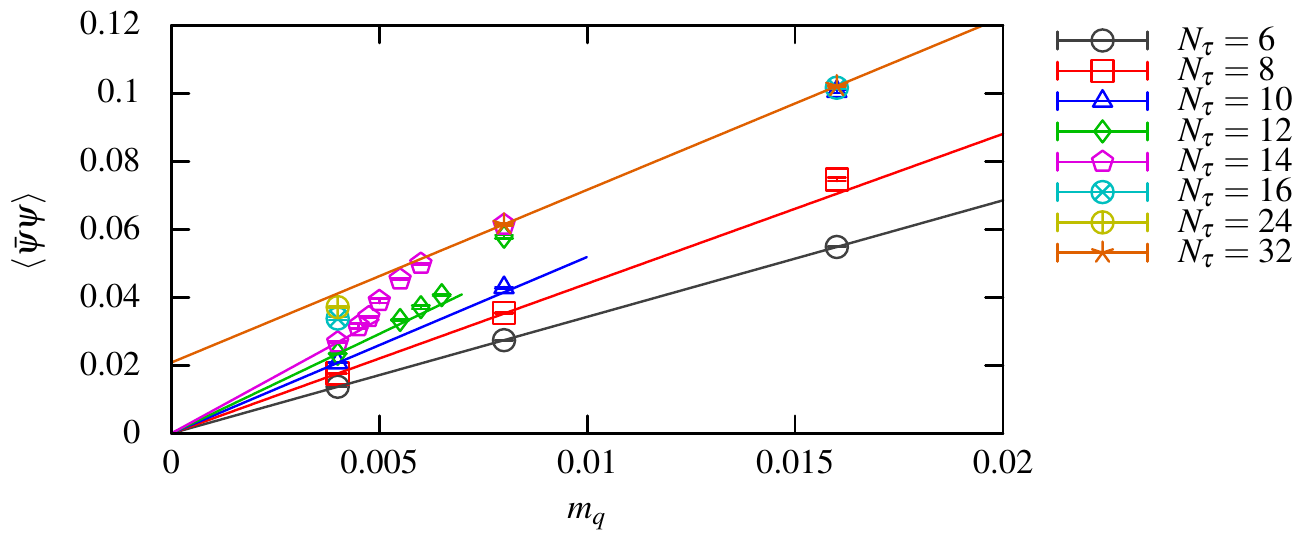}
  \caption{The values of $\pbp$ versus $m_q$ for 8 flavors at
    $\beta=0.56$.  The first order transition point cannot be resolved
    numerically with our current simulations.}
  \label{fig:b056_pbp}
\end{figure}

\begin{figure}
  \centering
  \includegraphics{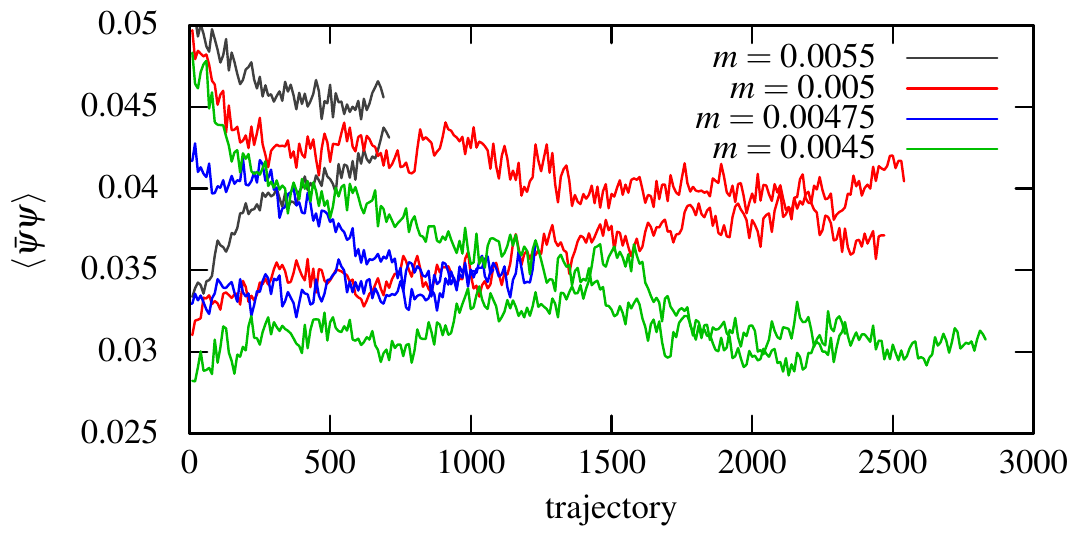}
  \caption{Evolutions of $\pbp$ with ordered and disordered starts at
    $N_\tau=14$, $\beta=0.56$.}
  \label{fig:b056_pbp_nt14}
\end{figure}

At weaker coupling, $\beta=0.56$, where the lattices are much finer
than those at $\beta=0.54$, we did not find a clear signal of a first
order transition.  Figure~\ref{fig:b056_pbp} shows a survey of the
parameter space we have covered.  Among those points, for
$m_q\le0.008$ and $N_\tau\ne32$, the spatial volume of the lattices
are $32^3$.  Similar to the previous figure, the line at $N_\tau=32$
is a linear extrapolation of $\pbp$ at zero temperature.  The lines
drawn through the origin denote the $m_q$ dependence of $\pbp$ at
finite temperature.  As we are interested in a possible discontinuity
of $\pbp$ we have done a careful scan, in quark mass, for $N_\tau =
14$.  Since we expect the discontinuity in $\pbp$ to be markedly
smaller here than for $\beta = 0.54$, we have focused on larger
$N_\tau$ and smaller quark masses.  Fixing $N_\tau=14$ and tuning the
quark mass, we could not find a clear metastability signal like the
ones seen at stronger coupling, $\beta=0.54$.  The evolutions of
$\pbp$ around $m_q=0.005$ at $N_\tau=14$ are shown in
Figure~\ref{fig:b056_pbp_nt14}.  In this figure, the evolution of
$\pbp$ at $m_q=0.005$ is very noisy, similar to the one in
Figure~\ref{fig:b054_pbp_nt6}, at $\beta=0.54$, $m_q=0.007$ with a
lattice size of $\lst{16}{6}$.  It is a good indication that we are
near the critical point of the transition, but either a much larger
lattice spatial volume is required to produce metastable states or the
quark mass is large enough so that the first order transition has
almost weakened to become second order.

\section{Conclusions}

We have found a clear two-state signal in lattice simulations of QCD
with 8 flavors at finite temperature, associated with the restoration
of chiral symmetry, on lattices where the chiral limit value for
$f_\pi = 0.06661(92)$.  The critical temperatures for two different
pion masses, in units of the pion decay constant, were determined and
are given in table~\ref{tab:tc_fpi}.  Extensive simulations at a
weaker coupling, and lighter quark masses, have shown that the
expected first order transition occurs for $N_\tau > 14$.  We have
some evidence that $N_\tau = 14$ is near the endpoint of the first
order transitions.  Much more computational time and careful tuning is
required to search for the first order signal at these weaker
couplings for 8 flavors.  Further work needs to be done for 12 flavors
to probe for the existence of first order thermal transition, which
would give more evidence for the phase of the 12 flavor system at zero
temperature.

We are thankful to all members of the RBC collaboration for useful
discussions, suggestions and help with the CPS software used in this
work.  Our calculations were done on the QCDOC and NY Blue at
BNL. This research utilized resources at the New York Center for
Computational Sciences at Stony Brook University/Brookhaven National
Laboratory which is supported by the U.S. Department, of Energy under
Contract No. DE-FG02-92ER40699 and by the, State of New York.

\bibliographystyle{JHEP-2}
\bibliography{ref}

\end{document}